# SRMAC - Smoothed Recursive Moving Average Crossover for Real-Time Systolic Peak Detection in Photoplethysmography

Cesar Abascal Machado • Victor O. Costa • Cesar Augusto Prior • Cesar Ramos Rodrigues

Universidade Federal de Santa Maria, Santa Maria, Brasil

**Abstract**

**Purpose.** Photoplethysmography (PPG) is a non-invasive technique that measures changes in blood flow volume through optical means. Accurate detection of peaks in the PPG waveform allows the extraction of the heart rate variability (HRV), a recognized indicator for the health of the cardiac and autonomic nervous systems. Previous research has established the feasibility of PPG peak detection based on the crossover of moving averages. This paper proposes the Smoothed Recursive Moving Average Crossover, which eliminates the need for post-processing and nonlinear pre-processing of previous crossover-based peak detectors. The proposed model is advantageous regarding memory and computational complexity, making it attractive for implementations on embedded devices.

**Methods.** Along with this paper, we make available a novel dataset comprising 66 minutes of PPG recordings from healthy individuals and patients with cardiopulmonary diseases, in three distinct phases of an acquisition protocol, with annotations for every systolic peak present. The optimization and assessment of the proposed peak detection model use this dataset. Its optimization is accomplished with the simple random search heuristic, while the leave-subject-out cross-validation method provides the means to assess its performance. The source code for all experiments reported in this research is also available in an online repository.

**Results.** The experimental study examines the performance of the proposed model considering different arrangements of the PPG data. These arrangements contemplate performances for each individual, for groups composed of healthy subjects and patients, and for the distinct acquisition protocol phases. The experiments show that the proposed model performs better than the previous crossover-based approach from the literature regarding the precision and recall metrics. More specifically, our model has an average precision of 0.9937 and an average recall of 0.9968.

**Conclusion.** The contribution of this research to the scientific community and literature is twofold. The dataset we collected is open for any researcher, and we improve upon the leading edge on crossover-based PPG peak detection. This improvement comes in terms of performance metrics and computational cost.

## I Introduction

The photoplethysmography (PPG) technique uses light to estimate changes in the blood volume in the microvascular bed of tissue (Allen, 2007). Since 1936, researchers have been studying the principles and usefulness of the PPG signal (Allen, 2007). After the introduction of pulse oximeters by Aoyagi et al. (1974), its importance for clinical medicine has increased (Nara et al., 2014). The basic configurations which enable PPG acquisition require only a light source and a photoreceptor, as shown in Figure 1 for different sensor placement configurations.

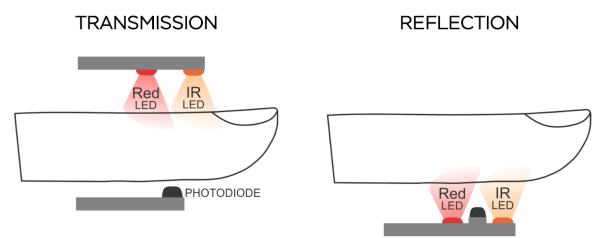

**Fig. 1** Common positioning schemes for light source and photoreceptors in fingertip PPG.

The PPG signal has AC and quasi-DC components. The pulsatile (AC) component of the signal is mainly influenced by arterial blood flow, with a weaker influence of venous blood, while the quasi-DC component relates to the interaction of light with tissues (Mohan et al. 2016). Systolic peaks, dicrotic notches and diastolic peaks are some of the patterns present in the PPG waveform (Park et al., 2022), as visible in Figure 2. The dicrotic notch is a small descending deflection in the diastolic phase, resulting from the closing of the aortic valve (Antonelli et al. 1994), and it tends to become more softened with advancing age due to the increasing stiffness of arteries through time (Allen & Murray, 2003).

A dual wavelength PPG recording can provide diverse physiological information, including blood oxygen saturation, heart rate, respiratory rate, blood pressure (Elgendi et al. 2018) and blood glucose levels (Habbu et al. 2019). Heart rate variability (HRV) is a time series consisting of intervals between heartbeats, and it is a valuable source of information on an individual's health status (Acharya et al., 2006). Traditionally, the HRV computation uses the R wave from electrocardiogram (ECG) signals. Due to the high correlation between R-R intervals and PPG peak-to-peak intervals, the HRV can be reliably estimated from PPG signals (Selvaraj et al. 2008). Both of these approaches use the outputs of peak detection models as a way to compute the HRV.

Given this correlation, the main advantages of PPG over ECG for HRV extraction are its simplicity and being less invasive. The process of ECG acquisition requires a trained professional to place electrodes on the subject's skin, and, in many cases, a preparation step includes hair removal and possibly conductive gel application. Another difficulty is that sources of electromagnetic noise, such as power line interference, may easily distort the ECG signal (Ziarani & Konrad, 2002). In such a way, researchers have focused on creating accurate algorithms for systolic peak detection in PPG signals to embed in various devices. Previous research has established the feasibility of models based on the crossover of moving averages for this objective (Elgendi et al., 2013) (Elgendi, 2016), which are both accurate and lightweight.

Although systolic peak detection in PPG may be affected by motion artifacts, it is possible to use an accelerometer sensor to discard unreliable portions of the signal from the HRV computation (Morelli et al., 2018). Moreover, the presence of dicrotic notches of different shapes may induce errors in peak detection and therefore algorithms should be able to deal with them.

The contribution of this study is twofold. 1) The proposal of a model for peak detection which further explores the crossover of moving averages. The Smoothed Recursive Moving Average Crossover (SRMAC) model uses lighter recursive moving averages and eliminates the need for both post-processing and nonlinear preprocessing present in previous crossover-based models and its low computational and memory complexities make it suitable for embedded real-time applications. 2) We make the complete dataset used in the experimental study available to the scientific community. The dataset comprises signals from 22 subjects, both healthy individuals and patients suffering from cardiopulmonary diseases, with an equal number of male and female individuals in the dataset. Each subject executed a three-stage protocol for signal recording. The first stage was at rest, the second one involved walking on a treadmill, and the last one was their recovery after physical exercise.

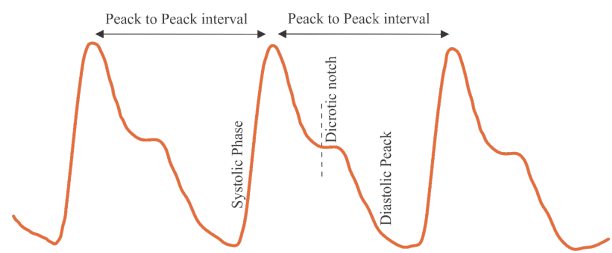

**Fig. 2** Typical photoplethysmography waveform.

The paper is organized as follows. Section II details the organization, acquisition and annotation of the provided dataset. In Section III, we describe the mechanisms underlying crossover-based peak detection and present the proposed model, along with the methodology used to optimize its parameters. Section IV presents the experimental study which validates the performance of SRMAC. Finally, Section V outlines the conclusions of the research. All source code necessary to implement SRMAC and replicate the experimental study is available in a

Table I - Statistical information about the volunteers

| Groups | | Age (y) | Weight (Kg) | Height (cm) | BMI (kg/m³) | Beats for one minute of acquisition | | | | | |
|---|---|---|---|---|---|---|---|---|---|---|---|
| | | | | | | Rest (4th min) | | Balke (12th min) | | Recovery (1st min) | |
| Healthy | Male (7) | 51.57±5.19 | 89.29±11.48 | 179.43±6.63 | 27.65±2.31 | 78.57±12.09 | 550 | 125.57±12.12 | 879 | 98.43±18.82 | 689 |
| | Female (4) | 49.50±2.38 | 69.00±15.55 | 162.25±8.66 | 26.02±4.41 | 80.50±6.76 | 322 | 115.75±11.00 | 463 | 90.75±97.10 | 363 |
| | All (11) | 50.82±4.35 | 81.91±16.01 | 173.18±11.13 | 27.05±3.12 | 79.27±10.12 | 872 | 122.00±12.21 | 1342 | 95.64±16.40 | 1052 |
| Not Healthy | Male (4) | 52.25±9.91 | 66.85±17.76 | 166.25±7.63 | 24.01±5.50 | 79.50±5.32 | 318 | 105.50±20.82 | 422 | 90.00±12.49 | 360 |
| | Female (7) | 60.43±6.35 | 67.83±13.09 | 160.71±7.93 | 26.04±3.20 | 85.86±12.31 | 601 | 123.14±16.75 | 862 | 106.29±20.21 | 744 |
| | All (11) | 57.45±8.41 | 67.47±14.06 | 162.73±7.94 | 25.30±4.03 | 83.55±10.47 | 919 | 116.73±19.43 | 1284 | 100.36±18.96 | 1104 |
| Total (22) | | 54.14±7.36 | 74.69±16.46 | 167.95±10.85 | 26.18±3.63 | 81.41±10.28 | 1791 | 119.36±16.06 | 2626 | 98.00±17.47 | 2156 |
| | | average ± standard deviation | | | | total | avg ± std | total | avg ± std | total | |

Git repository under the MIT license. The PPG peak detection dataset, consisting of waveforms and peak annotations, is available in a data repository.

## II Systolic Peak Detection Dataset

The dataset used in the optimization and validation of SRMAC comprises 66 minutes of PPG signals recorded at a sampling frequency of 200 Hz. The data comes from 22 volunteers, given that 11 of them are healthy individuals (seven men and four women, with a mean age of 50.82 ± 4.35 years and a mean body mass index (BMI) of 27.05 ± 3.12 Kg/m²) and the other 11 individuals have pulmonary diseases (four men and seven women, with a mean age of 57.45 ± 8.41 years and a mean BMI of 25.30 ± 4.03 Kg/m²). Healthy volunteers declared they do not have heart or lung problems, while unhealthy volunteers have been diagnosed with cardiopulmonary comorbidities such as asthma, chronic obstructive pulmonary disease, dysphagia, emphysema, hypertension and tuberculosis. Table I displays descriptive statistics of the subjects.

The authors informed all volunteers about the use of their data, and each volunteer filled out an Informed Consent Form (ICF) before the beginning of the experiment. The data collection approval by the ethics committee from the Federal University of Santa Maria (UFSM), Brazil, is registered under the number CAAE 63955616.5.0000.5346.

### II.A Acquisition of the PPG signal

A wireless portable PPG acquisition unit consisting of an AFE4490 analog front-end IC from Texas Instruments and a Raspberry Pi, model Zero W+ (Upton & Halfacree, 2014), was developed for this research. It employs a 22-bit analog-to-digital converter with drivers and time controls for PPG acquisition using 660 nm (red) and 940 nm (infrared) LEDs and sensors. Wireless communication controls the start and the end of recordings. A Mindray PPG sensor was integrated into the in-house equipment.

The proposed equipment was calibrated and tested against the Brain Wave II, manufactured by Neurovirtual and certified by ANVISA under registration number 80193710009.

### II.B Acquisition protocol

We developed a protocol to ensure all volunteers performed the same procedure during signal acquisition with the equipment described in the last subsection. Two physiotherapists from the University Hospital of Santa Maria (HUSM), Brazil, supervised the acquisition process to guarantee its correctness. This protocol, illustrated in Figure 3, instructed each volunteer to remain comfortably seated for 4 minutes, then to walk on a treadmill for 12 minutes under the Balke protocol (Meneghelo et al., 2010), and finally to sit comfortably to recover from the exercise for another 4 minutes.

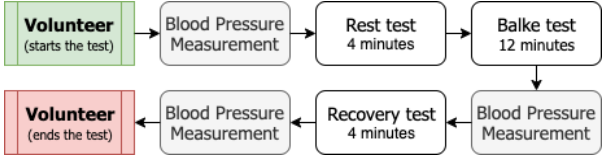

**Fig. 3** Test flowchart used during the acquisition of PPG signals.

The dataset captures the last minute of rest, the last minute of walking, and the first minute of recovery, producing signals in three different settings for each subject, resulting in 66 minutes of PPG signals. Furthermore, signals from the dataset we provide are available as raw data. Hence, they agree with those obtained in wearable systems, providing realistic conditions to evaluate the robustness of peak detection pipelines.

*II.C Dataset annotations*

All 66 1-minute samples were visualized graphically and had each systolic peak annotated by one of the authors of this study. Then, another author audited the samples to confirm the annotations. Altogether the dataset comprises 6,572 systolic peaks, 3265 of which are peaks for healthy volunteers and 3307 for unhealthy volunteers. The following subsection shows the file hierarchy of the dataset.

*II.D Dataset organization*

The dataset of raw PPG signals comes in comma-separated value files (CSV). Figure 4 shows the structure of the dataset. The dataset is hierarchized into directories according to types of volunteers and protocol phases, whereas directories corresponding to each protocol phase keep CSV files of PPG signals and systolic peak annotations.

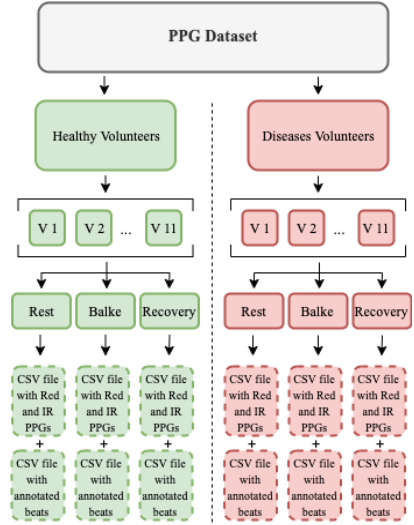

**Fig. 4** Structure of the photoplethysmography dataset.

## III Peak detection

The classical approach to peak detection in PPG signals consists of the manual inspection of waveforms and the annotation of peaks by experts. Although this strategy shows high hit rates, it is unfeasible for real-time applications.

Automatic peak detection techniques may rely on various principles, e.g. adaptive threshold (Shin et al., 2009), threshold based on past maximum values (Javier et al., 2021), transforms such as wavelets (Vadrevu and Manikandan, 2019), convolutional neural networks (Kazemi et al., 2022), and moving averages crossover models (Elgendi, 2016).

The following text deepens the discussion on crossover models and explains the proposed SRMAC model. Moving average crossover models exploit the fact that moving averages with distinct parameters have different delay characteristics. This family of models is employed in a diversity of areas, *e.g.* bearing failure detection (Phan and Tan, 2020) and technical analysis of financial markets (El-Khodary, 2009).

*III.A Moving average models*

Moving averages implement low-pass filters which smooths and delays an input signal. The delay of a moving average indicates how much it takes past samples into account for a given output sample.

Exponentially weighted moving averages (EWMA) are linear time-invariant (LTI) filters with infinite impulse response (IIR), and the following text shows how to derive it from simple moving averages (SMA), which are LTI finite impulse response (FIR) filters. The derivation below is based mainly on Prandoni and Vetterli (2008).

As a FIR filter, the output of a SMA can be computed with a convolution sum, as shown in equation 1 for an input signal $x$. The parameter $N$ determines how many samples the SMA takes into account, and therefore it determines the local average's smoothing power and the delay between its input and output signals. Using index substitutions and splitting the summation, equation 1 gives us equation 2. Furthermore, the definition of alpha from equation 3 leads to equation 4. For a large $N$, $y_{N-1}$ asymptotically approaches $y_N$, leading to the recursive forms of equations 5 and 6, in which $e$ is the output of the EWMA filter.

$$s_N[n] = \frac{1}{N} \sum_{k=0}^{N-1} x[n-k] \quad \text{(eq. 1)}$$

$$s_N[n] = \frac{N-1}{N} s_{N-1}[n-1] + \frac{1}{N} x[n] \quad \text{(eq. 2)}$$

$$\alpha = (N-1)/N \quad \text{(eq. 3)}$$

$$s_N[n] = \alpha\, s_{N-1}[n-1] + (1-\alpha)\, x[n] \quad \text{(eq. 4)}$$

$$s_N[n] \approx \alpha\, s_N[n-1] + (1-\alpha)\, x[n] \quad \text{(eq. 5)}$$

$$e[n] = \alpha\, e[n-1] + (1-\alpha)\, x[n] \quad \text{(eq. 6)}$$

$$h[n] = (1-\alpha)\, \alpha^n u[n] \quad \text{(eq. 7)}$$

Parameter α, valid in the interval [0, 1], governs the influence of older samples in the current output of the filter and therefore determines the delay and the smoothing power of the EWMA. As an LTI filter, the EWMA is fully described by its impulse response, shown in equation 7. For valid values of α, this impulse response is an exponential decay.

### III.B Crossover-based peak detection

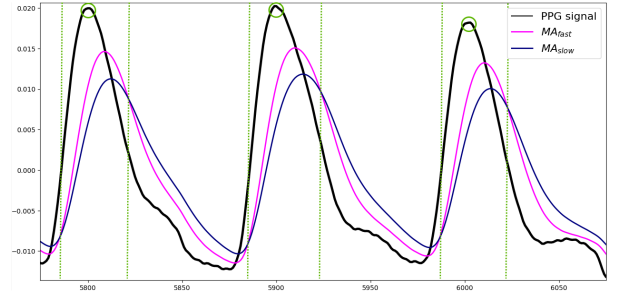

**Fig. 5** The outputs of two moving averages with distinct delays for a bandpass-filtered PPG signal as input. Green vertical lines indicate where moving averages cross each other and green circles indicate peaks.

Considering the moving averages with different parameters, $MA_{fast}$ and $MA_{slow}$, it is possible to build a system capable of peak detection for a given signal by subtracting the outputs of $MA_{slow}$ from $MA_{fast}$. The behavior that enables this capability is displayed in Figure 5 with an actual PPG signal, in which $MA_{fast}$ follows the signal faster than $MA_{slow}$ and surpasses it in the ascent of the PPG systolic peak, indicating the peak's beginning.

The regions of interest (ROI) for peaks can be found where the subtraction of averages trespasses a given threshold, which is a parameter of the detector, as shown in (in)equation 8. Ideally peaks should be found by searching for the maximum value of the input signal in each ROI.

$$MA_{fast}(x) - MA_{slow}(x) > thr \quad \text{(eq. 8)}$$

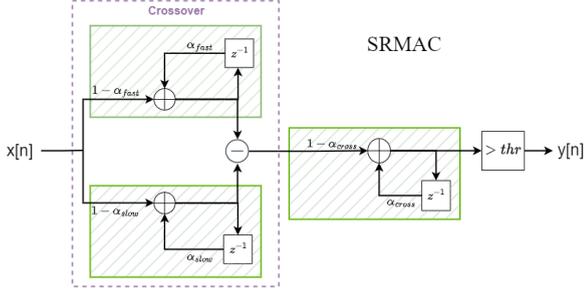

**Fig. 6** Representation of the input-output relationship of SRMAC. Green areas indicate the three EWMA of the model.

Evidence from the literature (Elgendi et al., 2013) suggests that, in the case of PPG peak detection, the crossover of two moving averages alone generates several spurious ROI, which contain no actual systolic peaks and end up evaluated as false positives. The crossover-based model we propose therefore filters the signal that results from subtraction using a third moving average, intending to smooth out short, spurious, ROI. Figure 6 displays a diagram for the proposed model, SRMAC, with discrete signals *x* and *y* as input and output, respectively. The EWMA filters are indicated in green boxes, and behave according to equation 6. In this way, SRMAC's mechanism to find ROI follows (in)equation 9.

$$e_{cross}(e_{fast}(x) - e_{slow}(x)) > thr \quad \text{(eq. 9)}$$

The SRMAC model has four distinct parameters: $\alpha_{fast}$, $\alpha_{slow}$, $\alpha_{cross}$ and threshold (*thr*). While parameters $\alpha_{fast}$ and $\alpha_{slow}$ govern the crossover dynamics, $\alpha_{cross}$ refers to the EWMA that filters spurious peaks before thresholding.

The signal processing pipeline used by (Elgend et al., 2013) also applies the crossover principle for the detection of systolic peaks in PPG signals. The Two Event-Related Moving Averages (TERMA) framework expanded the application of this pipeline to multiple types of biomedical signals (Elgendi, 2016), obtaining state-of-the-art results for many of them. Authors claim their framework is the first approach to use moving averages as decision-makers for biomedical signal processing. The pipeline from (Elgendi et al., 2013) comprises five steps.

Initially, there is a filtering of the input signal by a zero-phase bandpass filter, then the clipping of the resulting signal, *i.e.* truncation of its negative-valued samples to zero, and its squaring. The resulting signal feeds a crossover of SMA, which indicates ROI in the signal. A post-processing step eliminates short peaks to filter out spurious ROI generated by the crossover.

The number of samples that the SMA filter takes into account at each step, defined by the summation limits in equation 1, depends linearly on its parameter *N*, *i.e.* both computational and memory complexities w.r.t. *N* are O(*N*). In the EWMA filter, instead, computational and memory complexities are constant w.r.t. its parameter α, *i.e.* O(*1*). These observations are pertinent for implementations on embedded devices, which have constraints on available power and memory capacity. For implementations on dedicated hardware, it means that the SMA implementation will lead to a different design for each value of *N*, while the same EWMA implementation could deal with the entire domain of α.

### III.C Optimization of crossover models

To perform peak detection correctly, the models under analysis must undergo parameter optimization according to some performance metric. In this study, a systolic peak detected at time $t_{pred}$ counts as a true positive if there is an actual annotated peak at time *t* such that $|t - t_{pred}| < 0.1$ s. In this way, we may define true positives (TP) as the number of correctly detected peaks, false positives (FP) as the number of peak detections in the absence of actual peaks and false negatives (FN) as the number of true peaks which were not detected.

The literature on peak detection generally does not consider true negatives and can only use a subset of the metrics derived from confusion matrices. We follow this trend and use the precision (*Pp*) and recall (*SE*) metrics shown in equations 7 and 8. A high *Pp* indicates that the model rarely detects non-existent peaks, while a high *SE* is evidence of a model that rarely misses a peak.

$$Pp = \frac{TP}{TP + FP} \quad \text{(eq. 7)}$$

$$SE = \frac{TP}{TP + FN} \quad \text{(eq. 8)}$$

The experimental study in Section IV applies derivative-free heuristic optimization methods. During the optimization of a peak detection model, these methods generate sets of possible solutions and evaluate the performance of each one. The output of the optimization process is the best-performing solution found. In this study, we use the average between *Pp* and *SE* as the performance metric, or fitness function, and call it "accuracy".

Authors of TERMA report that they select parameters of the model with a brute force approach. We replicate this brute-force approach with a grid search heuristic, which, given a list of possible values for each parameter, evaluates all possible parameter combinations.

For the optimization of SRMAC, we use the random search heuristic. Random search samples sets of parameters from a continuous uniform distribution and evaluates the fitness of each solution. It is considered more efficient than grid search in hyperparameter optimization (Bergstra and Bengio, 2012).

## IV Experimental study

All experiments detailed in this section were executed using Python 3.9.13 with the NumPy module v. 1.22.4 (Harris et al., 2006) for vector math, the SciPy module v. 1.7.1 (Virtanen, 2020) for an implementation of Butterworth filters and the Matplotlib module v. 3.5.1 (Hunter, 2007) for some of the plots. The execution environment comprises an Intel® i7-1165G7 processor and 8 GB of RAM running the Windows 10 operating system.

### IV.A Description of experiments

This subsection describes the methodologies used to validate the proposed peak detector optimization. Table I summarizes statistics from the dataset. We use 66 minutes of raw infrared PPG recordings from 22 subjects, 11 of whom are COPD patients. Before peak detectors process the signals, a second-order Butterworth filters these.

According to its authors and the best of our knowledge, the TERMA framework is currently the only method in the literature to apply the crossover principle for systolic peak detection. Therefore, it serves as a basis for comparison to the proposed peak detector. The signal processing pipeline to which we compare our results is presented in (Elgendi et al., 2013) and briefly described in Section III.B of this manuscript.

The need for cross-validation (CV) comes from the fact that, due to the stochastic nature of sampling, splitting the collected dataset into train and test partitions could introduce biases to the results of our experiments. The CV approach instead splits a dataset into *k* different parts, or folds, and proceeds to optimize the model under evaluation *k* times. Each of these times, the optimization algorithm uses *k-1* folds of the dataset to train the model and reserves one of the folds, the validation fold, to evaluate the model obtained during optimization. Therefore, this procedure results in *k* sets of model parameters and their respective metrics on validation data.

The results we present for the models under evaluation employ the leave-subject-out cross-validation (LSOCV) strategy. This CV scheme sets the number of folds, *k*, to the number of individuals in the dataset. In this technique, the validation fold for each train-validation split comprises all observations gathered from a single subject. Therefore, performance evaluation of models only considers data from subjects it did not encounter in the optimization phase, thus preserving the within-subject correlation (Xu and Huang, 2012).

Given the stochasticity of the random search heuristic, statistics of each cross-validation round of training and validation incorporates results from 30 runs. The evaluation of grid search, a deterministic optimization method, does not require such repetitions. TERMA's optimization with grid search evaluates 1331 distinct combinations of parameters, including the best set of parameters reported in (Elgendi et al., 2013) and (Elgendi et al., 2016).

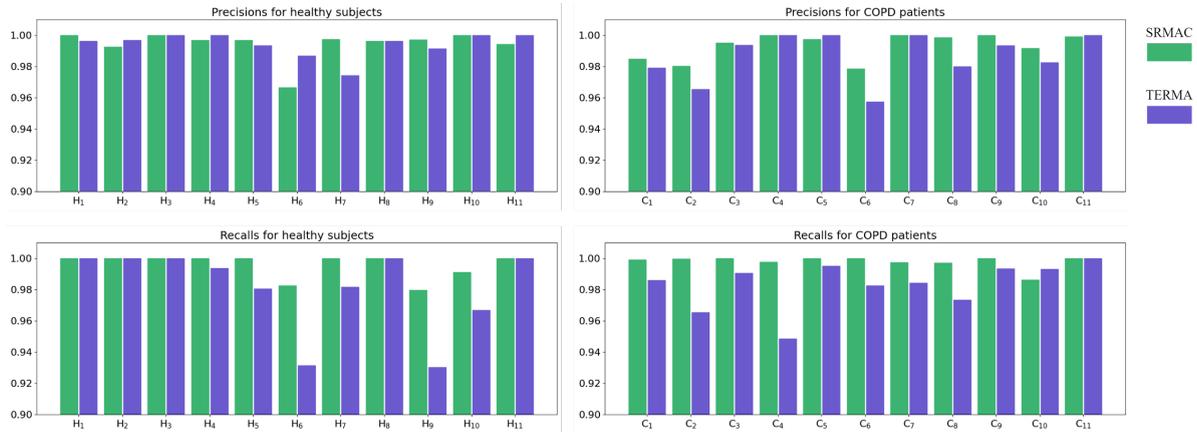

**Fig. 7** Performance metrics from individual subjects for both the proposed peak detector and TERMA.

Both grid and random search algorithms deal with constrained optimization problems. For the proposed peak detector, we optimize $\alpha_{fast}$, $\alpha_{slow}$ and $\alpha_{cross}$ on the interval $[0.7, 1)$ and the threshold on the interval $[0, 5E\text{–}4]$. The TERMA model has parameters $W_1$, $W_2$ and $\beta$, which are optimized in the intervals $[51, 111]$, $[545, 695]$ and $[0, 0.1]$, respectively. Parameters $W_1$ and $W_2$ are in ms, while alphas are dimensionless. Since the optimal cutoff frequencies for a bandpass filter depend on the intrinsic characteristics of the signal under analysis, we take the optimal cutoff frequencies reported in (Elgendi et al., 2013), 0.5 and 8 Hz, as prior knowledge and use them for all experiments.

### IV.B Results and discussion

Considering all 22 subjects from LSOCV, the overall accuracies of SRMAC and TERMA are 0.99528 and 0.98594, respectively. Table II breaks these metrics down into precisions and recalls for two types of subjects: healthy individuals and COPD patients. Although the average precision and recall values of SRMAC for healthy subjects and COPD patients are greater than TERMA's, the difference of precisions for healthy subjects seems negligible. For both methods, we observe that precision values of the healthy group are greater than precisions of the COPD group, while recalls for the COPD group are greater than recall values for the healthy one.

**Table II** - Average statistics for both SRMAC and TERMA, with standard deviations in parentheses.

| Model | Patients | Precision | Recall | Accuracy |
|---|---|---|---|---|
| **TERMA** | Healthy | 0.99408 (0.00781) | 0.98035 (0.02687) | 0.98722 |
| | COPD | 0.98646 (0.01497) | 0.98286 (0.01517) | 0.98466 |
| **SRMAC** | Healthy | 0.99431 (0.00951) | 0.99575 (0.00774) | 0.99503 |
| | COPD | 0.99316 (0.00825) | 0.99790 (0.00409) | 0.99553 |

As shown in Figure 7, the overall superiority of SRMAC does not mean that it performs better than TERMA for all subjects w.r.t. all performance metrics. For some subjects, such as the $C_{11}$ individual, TERMA may perform better or equally well when compared to SRMAC. Other subjects present particular difficulties to the performance of both methods. For the $H_6$ subject, for example, SRMAC and TERMA have lower precisions and recalls when compared to the average statistics of both models.

The distinct phases of the data acquisition protocol present different challenges to SRMAC. Therefore, we evaluate the performances of all parameter sets found during LSOCV for each protocol phase separately. Table III shows that both precision and recall in the "Walking" phase are lower than in "Rest" and "Recovery", which is evidence that the model 1) detects a few movement artifacts in the PPG signal as peaks and 2) misses a small number of peaks due to these artifacts. For an embedded implementation of SRMAC, one could use measures from an accelerometer to discard the few moments in which the PPG signal becomes unreliable enough to make the detection less accurate, as indicated in (Morelli et al., 2018).

Table III - Metrics by protocol phase.

| Phase | Precision | Recall | Accuracy |
|---|---|---|---|
| Rest | 0.99575 (0.00163) | 0.99959 (0.00098) | 0.99767 |
| Walking | 0.99258 (0.00115) | 0.99400 (0.00207) | 0.99329 |
| Recovery | 0.99586 (0.00063) | 0.99999 (8.175 E-5) | 0.99793 |

Table IV presents the overall average of validation accuracies for SRMAC throughout the search history, i.e. for different numbers of objective function evaluations (OFE). Although the train accuracy increases monotonically during the search, it is visible that the validation accuracy reaches its maximum value around 100 OFE. It is a symptom of overfitting, in which a decrease in the train error does not imply that the validation error will decrease, resulting from a lack of generalization by the optimized model. We then conclude that the random search heuristic with a few iterations is sufficient to reach the maximum performance of SRMAC.

Although the authors of TERMA optimize it with a brute force approach, we attempt its optimization with random search and also report its performance in Table IV. For the maximum number of OFE, it is evident that TERMA still did not reach its best possible performance, given that grid search has a greater overall accuracy. The reason for this is enlightened when we consider the best parameter sets found with grid search: for all LSOCV splits of our dataset, the best performing solutions have the parameter β equal to zero, thus eliminating the influence of this parameter in TERMA's peak detection threshold. It happens due to the inter- and intra-session variations in PPG signal magnitude of the dataset. With random search, which samples from a continuous uniform distribution, the probability of sampling equal to zero does not exist, or infinitesimally small if we consider the finite arithmetic precision of digital computers.

Table IV - Average LSOCV validation accuracies for different moments of the random search.

| OFE | SRMAC | TERMA |
|---|---|---|
| 50 | 0.99529 | 0.96750 |
| 100 | **0.99538** | 0.97551 |
| 150 | 0.99537 | 0.97828 |
| 200 | 0.99535 | 0.97978 |
| 250 | 0.99532 | 0.98028 |
| 300 | 0.99528 | **0.98091** |

When implementing peak detectors in embedded devices, one must account for the computational and memory requirements of the algorithms under consideration, along with the accuracies. Beyond using EWMA filters instead of SMA, the proposed solution to crossover-based peak detection improves the current state-of-the-art by eliminating all non-linear operations besides thresholding and replacing the post-processing step with a third EWMA. A future embedded implementation of the proposed SRMAC detector might demonstrate how much energy the aforementioned choices can save.

## V Conclusions

The contributions of the present research to the literature on systolic peak detection for PPG signals is twofold. We make available a peak-annotated dataset of PPG signals, with 66 minutes of recordings and a total of 6,573 systolic peaks, to the scientific community. The data extraction used three different protocol phases from 22 subjects, half of which are COPD patients, comprising a variety of heights, weights and ages. The second contribution is the proposal of SRMAC, a crossover-based peak detector that improves upon the state-of-the-art in terms of precision, recall and computational and memory complexities. Through an overfitting analysis, we demonstrate that the simple random search heuristic is sufficient to optimize SRMAC's four parameters.

## VI Acknowledgements

This study was financed in part by the Coordenação de Aperfeiçoamento de Pessoal de Nível Superior - Brasil (CAPES/PROEX) - Finance Code 001. The authors would also like to thank the Technology Center (CT) and the University Hospital of Santa Maria (HUSM), both from the Federal University of Santa Maria (UFSM), for the support provided during this research.

## VII References

Acharya, U. R., Joseph, K. P., Kannathal, N., Lim, C. M., & Suri, J. S. (2006). Heart rate variability: a review. *Medical and biological engineering and computing*, 44(12), 1031-1051.

Allen, J., & Murray, A. (2003). Age-related changes in the characteristics of the photoplethysmographic pulse shape at various body sites. *Physiological measurement*, 24(2), 297.

Allen J. (2007). Photoplethysmography and its application in clinical physiological measurement. *Physiological measurement*, 28(3), R1–R39.

Antonelli, L., Ohley, W. & Khamlach, R. (1994). Dicrotic notch detection using wavelet transform analysis. *Proceedings of 16th Annual International Conference of the IEEE Engineering in Medicine and Biology Society*. 216-1217.

Aoyagi, T., Kiahi, M., Yamaguchi, K., & Watanabe, S. (1974), Improvement of the earpiece oximeter. *Abstracts of the 13th Annual Meeting of the Japanese Society of Medical Electronics and Biological Engineering.*

Bergstra, J., & Bengio, Y. (2012). Random search for hyper-parameter optimization. *Journal of machine learning research*, 13(10), 281-305.

El-Khodary, I. A. (2009). A decision support system for technical analysis of financial markets based on the moving average crossover. *World Applied Sciences Journal*, 6(11), 1457-1472.

Elgendi, M., Liang, Y., & Ward, R. (2018). Toward generating more diagnostic features from photoplethysmogram waveforms. *Diseases*, 6(1), 20.

Elgendi, M., Norton, I., Brearley, M., Abbott, D., & Schuurmans, D. (2013). Systolic peak detection in acceleration photoplethysmograms measured from emergency responders in tropical conditions. *PLoS One*, 8(10), e76585.

Elgendi, M. (2016). TERMA framework for biomedical signal analysis: An economic-inspired approach. *Biosensors*, 6(4), 55.

Habbu, S. K., Joshi, S., Dale, M., & Ghongade, R. B. (2019). Noninvasive Blood Glucose Estimation Using Pulse Based Cepstral Coefficients. *2nd International Conference on Signal Processing and Information Security* (ICSPIS), IEEE. 1–4. doi:10.1109/ICSPIS48135.2019.9045897

Hunter, J. D. (2007). Matplotlib: A 2D Graphics Environment. *Computing in Science & Engineering*, 9(3), 90-95.

Harris, C. R., Millman, K. J., van der Walt, S. J. et al. (2006). Array programming with NumPy. *Nature*, 585, 357–362.

Instruments, T. (2014). AFE4490 Integrated analog front-end for pulse oximeters. Rev H.

Kazemi, K., Laitala, J., Azimi, I., Liljeberg, P., & Rahmani, A. M. (2022). Robust PPG Peak Detection Using Dilated Convolutional Neural Networks. *Sensors*, 22(16), 6054.

Kim, B. S., & Yoo, S. K. (2006). Motion artifact reduction in photoplethysmography using independent component analysis. *IEEE transactions on biomedical engineering*, 53(3), 566-568.

Meneghelo, R. S., Araújo, C. G. S., Stein, R., Mastrocolla, L. E., Albuquerque, P. F., & Serra, S. M. (2010). III Diretrizes da Sociedade Brasileira de Cardiologia sobre teste ergométrico. *Arquivos brasileiros de cardiologia*, 95(5), 1-26.

Mohan, P. M., Nisha, A. A., Nagarajan, V., & Jothi, E. S. J. (2016). Measurement of arterial oxygen saturation (SpO2) using PPG optical sensor. *Proceedings of the 2016 IEEE International Conference on Communication and Signal Processing.* 1136-1140. IEEE.

Morelli, D., Bartoloni, L., Colombo, M., Plans, D., & Clifton, D. A. (2018). Profiling the propagation of error from PPG to HRV features in a wearable physiological-monitoring device. *Healthcare technology letters*, 5(2), 59–64.

Nara, S., Kaur, M., & Verma, K. L. (2014). Novel notch detection algorithm for detection of dicrotic notch in PPG signals. *International Journal of Computer Applications*, 86(17), 36-39.

Park, J., Seok ,H. S., Kim S. & Shin H. (2022). Photoplethysmogram Analysis and Applications: An Integrative Review. *Frontiers in Physiology*. 12.

Phan, A.W.S., Tan, A.C. (2020). Bearing Failure Prediction Technique Using Exponential Moving Average Crossover Threshold with Support Vector Regression and Kernel Regression. *Proceedings of the 13th World Congress on Engineering Asset Management*. Springer. 667-675.

Prandoni, P., & Vetterli, M. (2008). *Signal processing for communications*. EPFL press.

Selvaraj, N., Jaryal, A., Santhosh, J., Deepak, K. K., & Anand, S. (2008). Assessment of heart rate variability derived from finger-tip photoplethysmography as compared to electrocardiography. *Journal of medical engineering & technology*, 32(6), 479-484.

Shelley, K. H. (2007). Photoplethysmography: beyond the calculation of arterial oxygen saturation and heart rate. *Anesthesia & Analgesia*, 105(6), S31-S36.

Shin, H. S., Lee, C., & Lee, M. (2009). Adaptive threshold method for the peak detection of photoplethysmographic waveform. *Computers in biology and medicine*, 39(12), 1145-1152.

Upton, E., & Halfacree, G. (2014). Raspberry Pi user guide. John Wiley & Sons

Vadrevu, S., & Manikandan, M. S. (2018). A robust pulse onset and peak detection method for automated PPG signal analysis system. *IEEE Transactions on Instrumentation and Measurement*, 68(3), 807-817.

Xu, G.; Huang, J. Z. (2012). Asymptotic optimality and efficient computation of the leave-subject-out cross-validation. *Annals of Statistics,* 40(6), 3003-3030.

Ziarani, A. K., & Konrad, A. (2002). A nonlinear adaptive method of elimination of power line interference in ECG signals. *IEEE transactions on biomedical engineering*, 49(6), 540-547.